\documentclass[sigconf]{acmart}
\usepackage{tabularx}

\AtBeginDocument{%
  }

\copyrightyear{2023}
\acmYear{2023}
\setcopyright{acmlicensed}\acmConference[WWW '23 Companion]{Companion Proceedings of the ACM Web Conference 2023}{April 30-May 4, 2023}{Austin, TX, USA}
\acmBooktitle{Companion Proceedings of the ACM Web Conference 2023 (WWW '23 Companion), April 30-May 4, 2023, Austin, TX, USA}
\acmPrice{15.00}
\acmDOI{10.1145/3543873.3587577}
\acmISBN{978-1-4503-9419-2/23/04}

\usepackage{subcaption}
\usepackage{multirow}
\usepackage{multicol}
\usepackage{algorithm}
\usepackage{algorithmic}
\usepackage{yang-macros}
\begin{document}

\title{Fairness-aware Differentially Private Collaborative Filtering}

\author{Zhenhuan Yang}
\email{zyang@etsy.com}
\affiliation{%
  \institution{Etsy Inc}
  \city{Brooklyn}
  \state{NY}
  \country{USA}
}

\author{Yingqiang Ge}
\email{yingqiang.ge@rutgers.edu}
\affiliation{%
  \institution{Rutgers University}
  \city{New Brunswick}
  \state{NJ}
  \country{USA}}
  
\author{Congzhe Su}
\email{csu@etsy.com}
\affiliation{%
  \institution{Etsy Inc}
  \city{Brooklyn}
  \state{NY}
  \country{USA}
}

\author{Dingxian Wang}
\email{dingxianwang@etsy.com}
\affiliation{%
  \institution{Etsy Inc}
  \city{Brooklyn}
  \state{NY}
  \country{USA}
}

\author{Xiaoting Zhao}
\email{xzhao@etsy.com}
\affiliation{%
  \institution{Etsy Inc}
  \city{Brooklyn}
  \state{NY}
  \country{USA}
}

\author{Yiming Ying}
\email{yying@albany.edu}
\affiliation{%
  \institution{University at Albany, SUNY}
  \city{Albany}
  \state{NY}
  \country{USA}
}

\renewcommand{\shortauthors}{Yang et al.}

\begin{abstract}
Recently, there has been an increasing adoption of differential privacy guided algorithms for privacy-preserving machine learning tasks. However, the use of such algorithms comes with trade-offs in terms of algorithmic fairness, which has been widely acknowledged. Specifically, we have empirically observed that the classical collaborative filtering method, trained by differentially private stochastic gradient descent (DP-SGD), results in a disparate impact on user groups with respect to different user engagement levels. This, in turn, causes the original unfair model to become even more biased against inactive users. To address the above issues, we propose \textbf{DP-Fair}, a two-stage framework for collaborative filtering based algorithms. Specifically, it combines differential privacy mechanisms with fairness constraints to protect user privacy while ensuring fair recommendations. The experimental results, based on Amazon datasets, and user history logs collected from Etsy, one of the largest e-commerce platforms, demonstrate that our proposed method exhibits superior performance in terms of both overall accuracy and user group fairness on both shallow and deep recommendation models compared to vanilla DP-SGD.


\end{abstract}



\maketitle

\section{Introduction}
With the explosive growth of e-commerce, consumers are increasingly relying on online platforms for their shopping needs. Traditional collaborative filtering (CF)-based recommendation models use a user's past interactions, such as ratings and clicks, to learn embeddings. However, the use of such user history can reveal sensitive information about the user. Prior works has shown an adversary can infer a targeted user’s actual ratings \cite{Hua2015DifferentiallyPM, liu2015fast} or deduce if the user is in the database \citep{berlioz2015applying, shin2018privacy} based on the recommendation list. To prevent privacy leakage, differential privacy (DP) \citep{dwork2006calibrating, dwork2014algorithmic} is a popular choice of mechanism in the above research as it provides theoretically quantitative privacy guarantee. In particular, differentially private stochastic gradient descent (DP-SGD) \citep{abadi2016deep, yang2021stability, yang2022differentially} is often adopted due to its scalability towards large neural networks.

\begin{figure*}[ht!]
\vspace{-5pt}
\captionsetup[subfloat]{labelformat=empty}
	\centering
	\small
		\subfloat[(a) Home \& Living]{\includegraphics[scale=0.235]{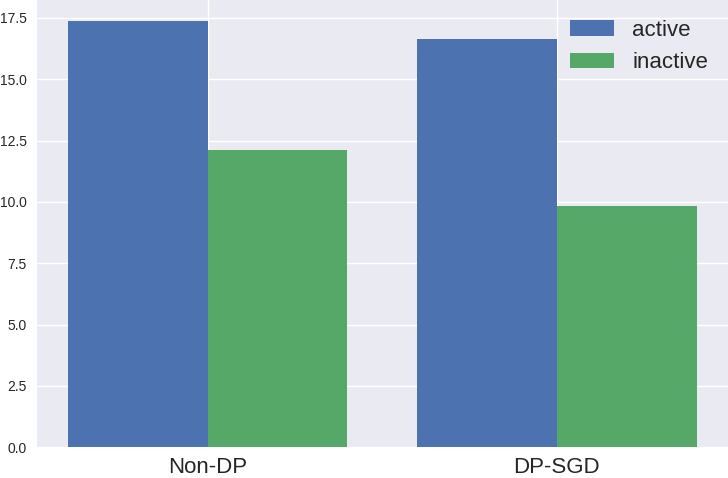}}\hfill
        \subfloat[(b) Craft Supplies \& Tools]{\includegraphics[scale=0.235]{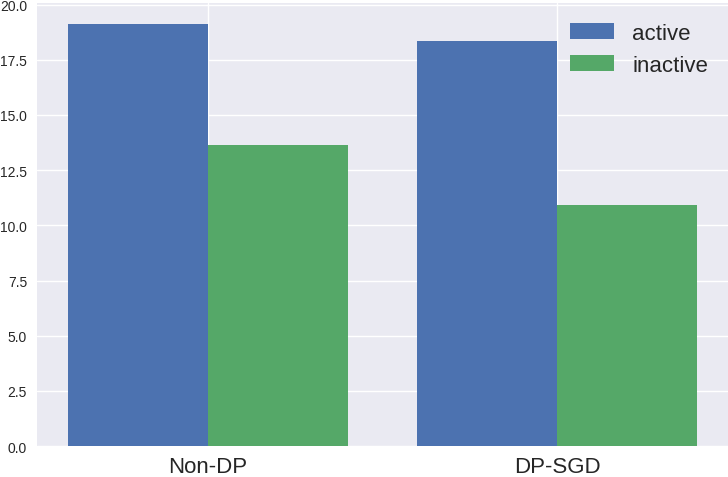}}\hfill
        \subfloat[(c) Grocery \& Gourmet Food]{\includegraphics[scale=0.235]{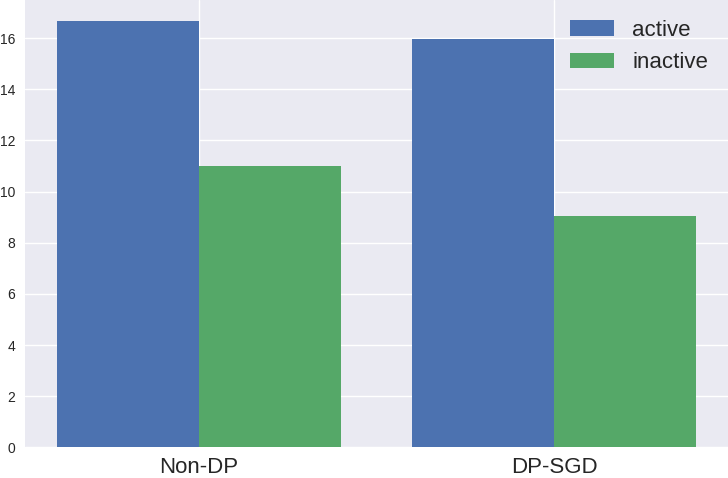}}\hfill
        \subfloat[(d) Beauty]{\includegraphics[scale=0.235]{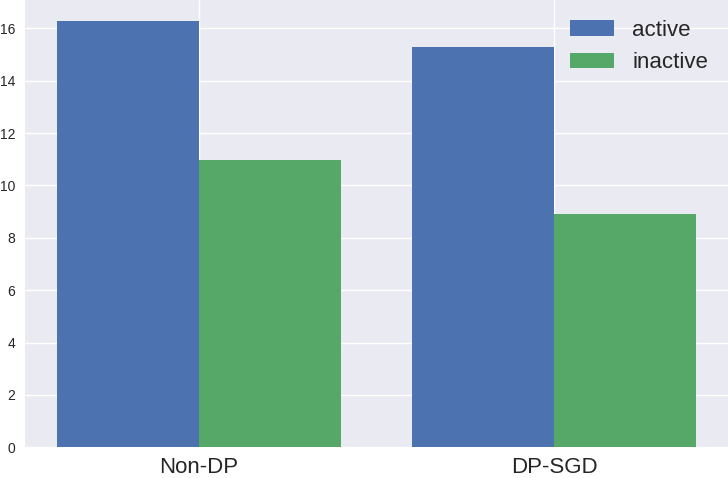}}\hfill
	\vspace{-2mm}	
  \caption{\small NDCG@10 (\%) on various datasets (See descriptions in Section \ref{sec:exp}) between active users (Blue) and inactive users (Green). In each subplot, left two bars labeled with Non-DP are NeuMF model trained by standard SGD. Right two bars are learned by DP-SGD with $\epsilon=1$. \label{fig:diff}}
\vspace{-10pt}
\end{figure*}

Despite the success of privacy protection by DP-SGD, it may also entail certain trade-offs. One such trade-off, recently discovered, is that the reduction in utility incurred by DP models disproportionately affects underrepresented subgroups in the image classification task \citep{bagdasaryan2019differential, bu2020deep}. In the field of recommendation systems, fairness concerns also arise due to other privacy protection mechanisms like federated learning \citep{liu2022fairness, maeng2022towards,zhu2022cali3f}. Therefore, it is natural to ask whether the protection of users' privacy will lead to disparate recommendation performance between different groups of users.

Specifically, in this work, we consider the unfair treatment between user groups with different activity levels, which is a common concern in the realm of fairness-aware recommendation \citep{zehlike2021fairness, ge2021towards, li2021user, li2021cikm, yang2022minimax}. In general, users who interact with the platform more frequently will contribute more sufficient data than those less active users when training the model. Due to the fundamental idea of collaborative filtering \cite{rendle2012bpr, he2017neural, 10.1145/3442381.3449864, ge2022survey}, it can lead to bias towards inactive users in the trained recommender system \citep{yao2017beyond, li2021user, fu2020fairness}. Consequently, users with lower activity levels are more likely to receive lower NDCG or so to speak unsatisfactory recommendations as demonstrated in Figure \ref{fig:diff} by Non-DP bars. What's worse, our empirical findings reveal that this accuracy bias worsens when differential privacy is introduced. Specifically, the NDCG of inactive users takes a heavier hit comparing with the active users, namely, the unfair treatment measured by the NDCG gap between active and inactive users in the DP-SGD setting increases by 15-20\% comparing with Non-DP setting across four datasets. This is also called ``the poor become poorer'' phenomenon (Matthew Effect) as in \cite{bagdasaryan2019differential, 10.1145/3397271.3401431}.

To this end, we propose a novel two-stage framework for CF-based approaches to address the issues of utility degradation and unfairness aggravation brought by DP-SGD, called \textbf{DP-Fair}. Specifically, the first stage of DP-Fair applies noise perturbation to user and item embeddings separately, thereby improving utility while providing privacy guarantees over vanilla DP-SGD. In the second stage, we apply a post-processing step to enforce user group fairness on the final recommendation list via solving of an integer programming problem. Our experimental results on Amazon benchmark datasets and user history logs collected on Etsy, demonstrate that our proposed algorithm outperforms vanilla DP-SGD based collaborative filtering in terms of overall recommendation performance and user-side group fairness.


\section{Preliminaries}

\subsection{Collaborative Filtering Models} Let $\Ucal = \{u_1, \cdots, u_{n_1}\}$ and $\Vcal =\{v_1, \cdots, v_{n_2}\}$ be the sets of users and items, respectively. Let $\Hcal_u = \{v\in \Vcal\}$ denote the set of items that user $u$ had positive interactions with. It is worth noting that we treat all interactions as binary implicit feedback (e.g. one if there is a click). Explicit feedback such as rating $r$ (e.g. 1-5) are converted to one if $r>3$ otherwise zero. $\Hcal$ denote the collection of all $\Hcal_u$. Let $n$ be the number of total positive interactions, i.e. $|\Hcal| = n$. Let $\xbf_u\in \Rbb^{n_1}, \xbf_v\in \Rbb^{n_2}$ be the one-hot encoding of the user $u$ and item $v$, respectively. Let $\zbf_u = U\xbf_u \in \Rbb^{d_1}, \zbf_v = V\xbf_v \in \Rbb^{d_2}$ denote the corresponding latent embeddings. We also employ $W$ to denote any other potential feature extraction parameters and $\Theta = (U,V,W)$ to denote all learnable parameters. 

A collaborative filtering latent factor model $f_\Theta: \Ucal \times \Vcal \rightarrow \Rbb$ is learning to infer the implicit feedback pattern once a learning to rank loss $l$ is given. In this work, we focus on the classic Bayesian Personalized Ranking (BPR) \citep{rendle2012bpr} loss, as follows, 
\begin{align*}
l(f_\Theta) =\! \sum_{u, v, v'} \!& -\log \sigma(f_\Theta(\xbf_u, \xbf_v) - f_\Theta(\xbf_u, \xbf_{v'})) + \frac{\lambda}{2}\|\Theta\|^2,
\end{align*}
where $\sigma$ is the sigmod function, $\lambda$ is the regularization parameter, $v \in \Hcal_u$, and $v' \in \Hcal_u^- = \Vcal \setminus\Hcal_u$ denotes an item that user $u$ does not provide implicit feedback. Since the positive interactions are usually more sparse, we slightly abuse the notation and let $\Hcal^-$ also denote a uniformly sub-sampled set of itself such that $|\Hcal^-| = |\Hcal| = n$. After learning, a recommendation list $\Rcal_u^k = \{v\in \Vcal\}$ for each user $u$ is produced based on top $k$ ranking scores $\{f_\Theta(u, \Vcal)\}$.

\subsection{Differential Privacy and DP-SGD}  

We first introduce the definition of differential privacy, which is given as follow.
\begin{definition}\label{def:differential-privacy}
For any $\epsilon, \delta > 0$, an (randomized) algorithm $\Acal$ is said to be $(\epsilon, \delta)$-differentially private if for all neighboring datasets $D,D'$ that differs by at most one example, and for all possible output sets $\Theta$ by $\Acal$, there holds
\begin{align*}
\Pbb[\Acal(D) \in \Theta] \leq \exp(\epsilon) \Pbb[\Acal(D') \in \Theta] + \delta.
\end{align*}
\end{definition}

\noindent where $\epsilon$ denotes the privacy budget (smaller values indicates a stronger privacy guarantee) and $\delta$ denotes tolerance of probability that the privacy guarantee fails. In practice, it often requires $\delta \ll \frac{1}{n}$. Since users' sensitive information can be inferred from the interaction data, the private dataset in this case is $D = \Hcal \cup \Hcal^-$.

At each iteration $t$, DP-SGD performs gradient norm clipping with some bound $C$ and Gaussian noise addition with variance $\sigma^2$ on the received gradients $G_t$, and then performs regular SGD on the model parameter $\Theta_t$ based on the new gradients $\tilde{G}_t$. If one randomly samples a batch $\Bcal_t \subseteq \Hcal$ of size $m$, then for each example $S \in \Bcal_t$, DP-SGD runs as
\begin{align*}
\bar{G}_t(S) = & G_t(S) / \max\{1, \|G_t(S)\|_2 / C\}\label{eq:gradient-clipping}    \numberthis\\
\tilde{G}_t(S) = & \bar{G}_t(S) + \Ncal(0, \sigma^2\Ibf) \label{eq:noise-addition}\numberthis
\end{align*}

\subsection{User-side Fairness} Let $\Rcal_u^K$ denote the recommendation list to user $u$ originally. We utilize the non-parity unfairness measure initially introduced in \citet{kamishima2011fairness}, stated as follow:
\begin{definition}\label{def:non-parity-unfairness}
Given a recommendation evaluation metric $\Mcal$, the use group fairness with respect to groups $A$ and $B$ is defined as 
\begin{align*}
\Ebb_u[\Mcal(\Rcal_u^K) | u\in A] = \Ebb_u [\Mcal(\Rcal_u^K) | u\in B].
\end{align*}

Empirically, the user group fairness is measured by
\begin{align*}
\mathcal{F}_{U}(\Rcal^K; A, B) = \Big|\frac{1}{|A|} \!\sum_{u \in A}\Mcal(\Rcal_u^K) \!-\! \frac{1}{|B|} \!\sum_{u \in B}\Mcal(\Rcal_u^K)\Big|.
\end{align*}
\end{definition}
Due to different user activity levels, recommender systems would usually underperform against users who have less historical interactions. This bias can be amplified when differential privacy is incorporated in the model as demonstrated in \cite{bagdasaryan2019differential}. Following the classical 80/20 rule, we select the top 20\% users based on their engagement activities as the frequent/active group $A$ and the rest as the infrequent/inactive group $B$.

\section{Fairness-aware Differentially Private Collaborative Filtering}

\begin{algorithm}[ht!]
\begin{algorithmic}[1]
\caption{\bf DP-Fair \label{alg:dp-sgd-new}}
\STATE {\bf Inputs:} Private dataset $\Dcal = \Hcal \cup \Hcal^-$; privacy parameters $\epsilon, \delta$; number of iterations $T$; learning rate $\{\eta_t: t \in [T]\}$, mini-batch size $m$; initial points $\Theta_0$. 
\STATE {\it \textbf{Stage I: Private Training}}
\FOR{$t=0$ to $T-1$}
\STATE Randomly sample a batch $\Bcal_t \subseteq \Dcal$ of size $m$
\FOR{each example $S \in \Bcal_t$} 
\STATE Apply DP-SGD steps as in Eq. \eqref{eq:gradient-clipping} and \eqref{eq:noise-addition} separately for user and item gradients $\tilde{G}_t(S) = (\tilde{G}_u(S), \tilde{G}_v(S))$
\ENDFOR
\STATE Update $\Theta_{t+1} = \Theta_t - \frac{\eta_t}{m} \sum_{S \in B_t}\tilde{G}_t(S)$
\ENDFOR
\STATE {\it \textbf{Stage II: Fairness Re-ranking}}
\STATE Rank based on $f_\Theta$ and return long recommendation lists $\Rcal^K$
\STATE Solve Eq. \eqref{eq:ugf-opt} via integer programming
\STATE {\bf Outputs:} Short new top recommendation lists $\Rcal^k$
\end{algorithmic}
\end{algorithm}

\begin{figure*}[t]
\vspace{-10pt}
	\centering
 \captionsetup[subfloat]{labelformat=empty}
	\small
		\subfloat{\includegraphics[scale=0.235]{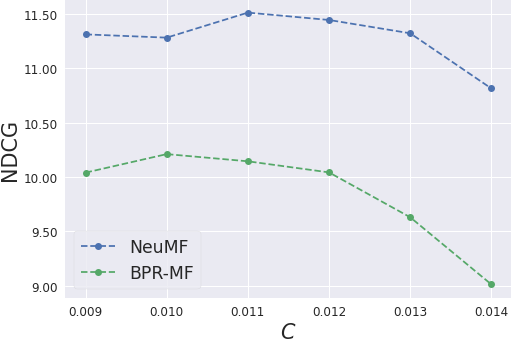}}\hfill
		\subfloat{\includegraphics[scale=0.235]{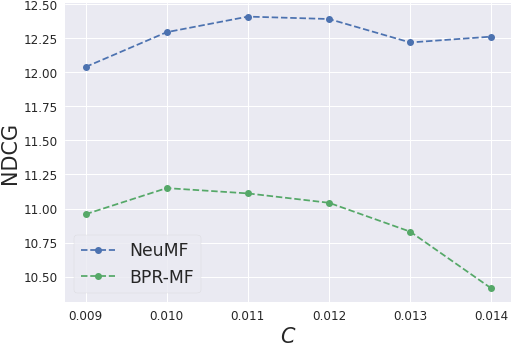}}\hfill
        \subfloat{\includegraphics[scale=0.235]{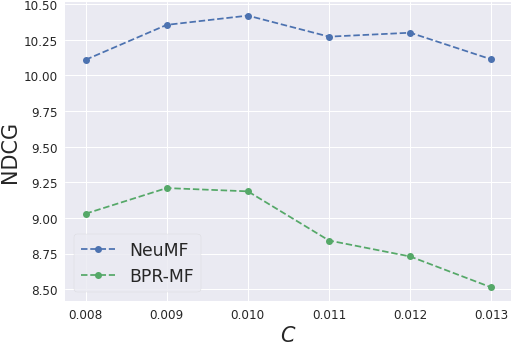}}\hfill
        \subfloat{\includegraphics[scale=0.235]{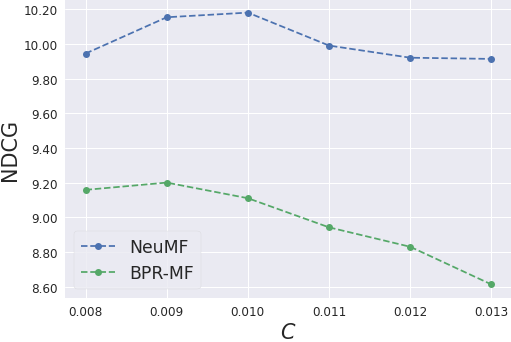}}\hfill
		\subfloat[(a) Home \& Living]{\includegraphics[scale=0.235]{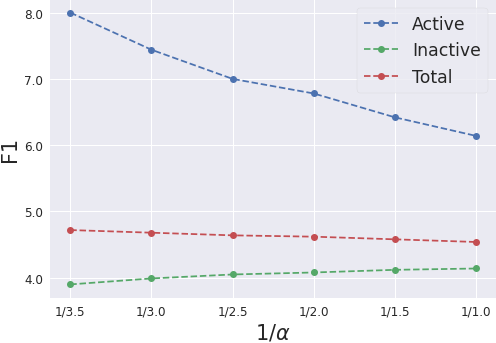}}\hfill
		\subfloat[(b) Craft Supplies \& Tools]{\includegraphics[scale=0.235]{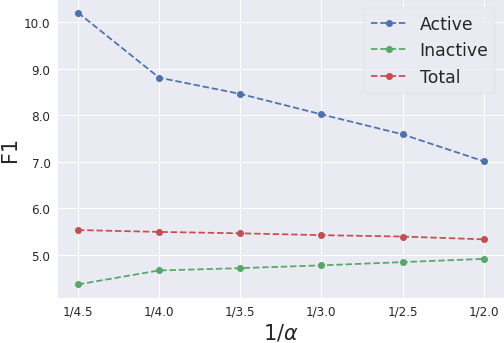}}\hfill
        \subfloat[(c) Grocery \& Gourmet Food]{\includegraphics[scale=0.235]{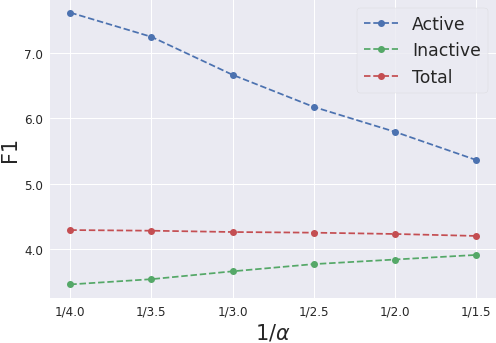}}\hfill
        \subfloat[(d) Beauty]{\includegraphics[scale=0.235]{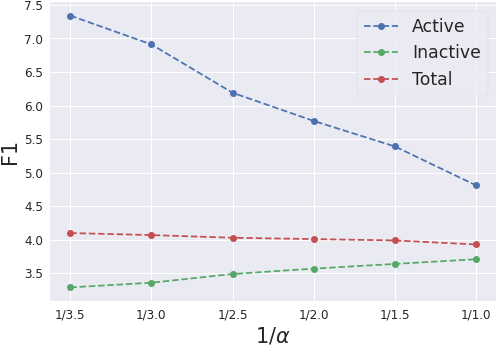}}
	\vspace{-2mm}
	\caption{\small Top: Performance results with respect to different values of clipping bound $C$ in terms of NCDG. Bottom: Performance results with respect to different levels of fairness constraint $\alpha$ in terms of F1.\label{fig:ablation}}
 \vspace{-10pt}
\end{figure*}

\noindent There are two stages in our proposed framework, DP-Fair. Given a private dataset $D$ with user provided privacy budgets $(\epsilon, \delta)$, the first stage applies DP-SGD for training the BPR and providing privacy guarantee. It is worth noting that at Line 6 we replace the uniform DP-SGD with separated ones on user and item. This procedure can avoid unnecessary norm clipping and noise addition \citep{Hua2015DifferentiallyPM,mcmahan2017learning} since user and item gradients may be different in scale during training. Overall, this tailored DP step will lead to better utility than vanilla DP-SGD with less perturbed gradients.

In the second stage, in order to mitigate the unfair treatment, we employ a post-processing approach \citep{li2021user}. At Line 12, once top-$K$ ranking lists $\Rcal^K$ are available, we re-rank them by maximizing the sum of prediction scores under the user group fairness constraint 
\begin{align*}\label{eq:ugf-opt}
\max_{\Rcal^k} \quad & \sum_{u \in \Ucal} \sum_{v \in \Rcal_u^k} f_\Theta(\xbf_u, \xbf_v) \numberthis \\
\textrm{s.t.} \quad & \mathcal{F}_{U}(\Rcal^k; A,B) \leq \alpha \text{ and } \Rcal_u^k \subseteq \Rcal_u^K \quad \forall u \in \Ucal.  
\end{align*}

One can consider $\Rcal^K$ as a binary matrix $R \in \Rbb^{n_1 \times K}$ where $R_{u,v} = 1$ means item $v$ is recommended to the user $u$. Hence the optimization problem \eqref{eq:ugf-opt} can be translated and solved as a $0-1$ integer programming problem. For ranking metric $\Mcal$, we pick the commonly used F1 score, which makes the computation more efficient than NDCG since it avoids the position discounted effect. It is worth noting that since this method is post-processing based on the recommendation list only, it will not break the differential privacy guarantee over the learned parameters $f_\Theta$ \citep{dwork2006calibrating}.

\section{Experiments\label{sec:exp}}
\subsection{Experimental Setup}
\subsubsection{\bf Datasets} We utilized two distinct sources of data. Firstly, we employ a benchmark dataset, namely the Amazon review dataset (5-core), which includes product reviews from the \textit{Grocery \& Gourmet Food} and \textit{Beauty} categories \citep{he2016ups}. Since both are encoded with explicit feedback through ratings, we transform them into binary feedback. Secondly, we collect and sample one month's worth of user history logs from two categories---\textit{Home \& Living} and \textit{Craft Supplies \& Tool}, on Etsy, one of the largest e-commerce platforms. We consider users' clicks as positive feedback in both datasets.

\begin{table}[h]
\setlength\tabcolsep{1.5pt}
    \centering
    \small
    \begin{tabular}{l|c|c|c|c}
    \toprule
      Name & \# User & \# Item & 
      \# Interactions & Sparsity \\
      \midrule
            Home \& Living & 6,538 & 3,924 & 100,855 & 99.61\% \\
            Craft Supplies \& Tool & 4,488 & 5,569 & 159,445 & 99.36\%\\
    Grocery \& Gourmet Food & 14,681 & 8,713 & 151,254 & 99.88\% \\
      Beauty & 22,363 & 12,101 & 198,502 & 99.93\% \\
    \bottomrule
    \end{tabular}
    \caption{Statistics of datasets.\label{tab:etsy-stats}}
\vspace{-30pt}
\end{table}

\subsubsection{\bf Implementation Details.} In this experiment, we performe a randomized 8:1:1 split of the datasets to create training, validation, and test sets.  We consider both shallow and deep recommendation models, namely BPR-MF \citep{rendle2012bpr} and NeuMF \citep{he2017neural}. To find the best clipping bounds, we follow the DP tuning strategy in \citet{mcmahan2018general} via pre-trainining. We set pre-ranker $K=20$ and re-ranker $k=10$. We fix the privacy parameter $\delta = \frac{1}{n^{1.5}}$ and employ the Opacus\footnote{\url{https://opacus.ai/}} module to conduct DP-SGD steps. We also employ the Gurobi\footnote{\url{https://www.gurobi.com}} solver to solve the re-ranking problem in Eq. \eqref{eq:ugf-opt}. 

\begin{table*}[ht!]
\setlength\tabcolsep{1.5pt}
\renewcommand{\arraystretch}{.6}
\centering\begin{tabularx}{\linewidth}{p{1.5cm}p{1.5cm}p{1.5cm}XXXXXXXXXXXX}
\toprule
\multirow{2}{*}{}& & &\multicolumn{4}{c}{$\epsilon=\infty$} & \multicolumn{4}{c}{$\epsilon=10$} & \multicolumn{4}{c}{$\epsilon=1$} \\ 
\cmidrule(lr){4-7}
\cmidrule(lr){8-11}
\cmidrule(lr){12-15}
Model & Metric & Algorithm & Total $\uparrow$ &Act. $\uparrow$ &InAct. $\uparrow$  &$\mathcal{F}_{U}$ $\downarrow$ &Total $\uparrow$ &Act. $\uparrow$  &InAct. $\uparrow$ & $\mathcal{F}_{U}$ $\downarrow$  &Total $\uparrow$  &Act. $\uparrow$ &InAct. $\uparrow$  & $\mathcal{F}_{U}$ $\downarrow$ \\
\midrule

\multicolumn{15}{c}{Home \& Living}\\
\midrule

\multirow{4}{*}{BPR-MF}&
\multirow{2}{*}{NDCG} &
DP-SGD &11.57 &15.17 &10.73 &4.44 &10.77 &14.93&9.67&5.25&10.21&14.44&9.15&5.29\\
&& DP-Fair &\textbf{12.04} &14.18 &\textbf{11.52} &\textbf{2.57} &\textbf{11.19}&14.09&\textbf{10.44}&\textbf{3.64}&\textbf{10.61}&13.55&\textbf{9.87}&\textbf{3.68}\\
&\multirow{2}{*}{F1}&
DP-SGD &4.85 &7.17 &4.39 &2.77 &4.56&6.96&3.90&2.97&4.31&6.84&3.67&3.17\\
&& DP-Fair &\textbf{4.89} &6.57 &\textbf{4.51} &\textbf{2.07} &\textbf{4.61}&6.38&\textbf{4.11}&\textbf{2.27}&\textbf{4.33}&6.30&\textbf{3.83}&\textbf{2.47}\\

\midrule

\multirow{4}{*}{NeuMF}&
\multirow{2}{*}{NDCG} &
DP-SGD &12.24 &17.36 &11.12 &6.21 &11.61&16.73&10.17&7.19&11.19&16.64&9.82&6.82\\
&& DP-Fair &\textbf{12.93} &16.12&\textbf{12.13} &\textbf{3.99} &\textbf{12.05}&15.78&\textbf{11.12}&\textbf{4.66}&\textbf{11.59}&15.01&\textbf{10.74}&\textbf{4.27}\\
&\multirow{2}{*}{F1}&
DP-SGD &5.03 &7.65 &4.37 &3.28 &4.78&7.47&4.10&3.36&4.58&7.61&3.82&3.79\\
&& DP-Fair &\textbf{5.08} &7.28 &\textbf{4.53} &\textbf{2.75} &\textbf{4.81}&7.20&\textbf{4.21}&\textbf{2.99}&\textbf{4.64}&6.96&\textbf{4.05}&\textbf{2.91}\\ 
\midrule
\multicolumn{15}{c}{Craft Supplies \& Tools}\\
\midrule

\multirow{4}{*}{BPR-MF}&
\multirow{2}{*}{NDCG} &
DP-SGD &12.63 &16.51 &11.72 &4.79 &11.76&16.28&10.57&5.72&11.15&15.75&10.00&5.75\\
&& DP-Fair &\textbf{13.59} &16.02 &\textbf{13.01} &\textbf{3.01} &\textbf{12.64}&15.91&\textbf{11.80}&\textbf{4.11}&\textbf{11.98}&15.32&\textbf{11.15}&\textbf{4.17}\\
&\multirow{2}{*}{F1}&
DP-SGD &5.30 &7.89 &4.80 &3.09 &4.99&7.30&4.27&3.61&4.71&7.48&4.02&3.45\\
&& DP-Fair &\textbf{5.36} &7.22 &\textbf{4.95} &\textbf{2.16} &\textbf{5.06}&7.01&\textbf{4.52}&\textbf{2.49}&\textbf{4.75}&6.91&\textbf{4.21}&\textbf{2.70}\\

\midrule

\multirow{4}{*}{NeuMF}&
\multirow{2}{*}{NDCG} &
DP-SGD &13.57 &19.12 &12.37 &6.74 &12.88&18.37&11.32&7.05&12.41&18.35&10.93&7.41\\
&& DP-Fair &\textbf{14.55} &17.76 &\textbf{13.65} &\textbf{4.11} &\textbf{13.56}&17.62&\textbf{12.51}&\textbf{5.11}&\textbf{13.16}&17.49&\textbf{12.08}&\textbf{5.40}\\
&\multirow{2}{*}{F1}&
DP-SGD &5.91 &9.01 &5.14 &3.86 &5.62&8.98&4.78&4.19&5.39&8.71&4.56&4.15\\
&& DP-Fair &\textbf{5.98} &8.55 &\textbf{5.33} &\textbf{3.21} &\textbf{5.64}&8.41&\textbf{4.95}&\textbf{3.46}&\textbf{5.46}&8.44&\textbf{4.71}&\textbf{3.73}\\ 

\midrule

\multicolumn{15}{c}{Grocery \& Gourmet Food}\\
\midrule

\multirow{4}{*}{BPR-MF}&
\multirow{2}{*}{NDCG} &
DP-SGD &10.65 &14.24 &9.80 &4.45 &9.90&14.04&8.81&5.23&9.21&13.57&8.12&5.45\\
&& DP-Fair &\textbf{11.34} &13.88 & \textbf{10.85} &\textbf{3.03} &\textbf{10.54}&13.27&\textbf{9.70}&\textbf{3.58}&\textbf{9.79}&12.64&\textbf{9.08}&\textbf{3.56}\\
&\multirow{2}{*}{F1}&
DP-SGD &4.22 &6.04 &3.79 &2.26 &4.03&5.92&3.52&2.39&3.78&5.74&3.28&2.46\\
&& DP-Fair &\textbf{4.28} &5.78 & \textbf{3.91} &\textbf{1.87} &\textbf{4.09}&5.71&\textbf{3.67}&\textbf{2.05}&\textbf{3.82}&5.44&\textbf{3.41}&\textbf{2.03}\\

\midrule

\multirow{4}{*}{NeuMF}&
\multirow{2}{*}{NDCG} &
DP-SGD &11.40 &16.66 &10.23 &6.43 &10.82&16.08&9.36&6.72&10.42&15.98&9.03&6.95\\
&& DP-Fair &\textbf{11.87} &15.31 &\textbf{11.01} &\textbf{4.29} &\textbf{11.06}&14.94&\textbf{10.09}&\textbf{4.85}&\textbf{10.73}&14.69&\textbf{9.74}&\textbf{4.95}\\
&\multirow{2}{*}{F1}&
DP-SGD &4.64 &7.01 &4.05 &2.96 &4.41&6.76&3.76&3.25&4.23&6.79&3.59&3.20\\
&& DP-Fair &\textbf{4.67} &6.66 &\textbf{4.23}&\textbf{2.43} &\textbf{4.42}&6.39&\textbf{3.87}&\textbf{2.52}&\textbf{4.26}&6.62&\textbf{3.66}&\textbf{2.96}\\ 

\midrule

\multicolumn{15}{c}{Beauty}\\
\midrule

\multirow{4}{*}{BPR-MF}&
\multirow{2}{*}{NDCG} &
DP-SGD &10.43 &13.66 &9.63 &4.02 &9.68&13.61&8.68&4.93&9.20&13.16&8.22&4.94\\
&& DP-Fair &\textbf{11.22} &13.15&\textbf{10.75} &\textbf{2.39}&\textbf{10.43}&13.13&\textbf{9.75}&\textbf{3.39}&\textbf{9.76}&12.20&\textbf{9.14}&\textbf{3.06}\\
&\multirow{2}{*}{F1}&
DP-SGD &4.02 &5.50 &3.65 &1.85 &3.82&5.62&3.37&2.25&3.54&5.38&3.08&2.29\\
&& DP-Fair &\textbf{4.10} &5.20 &\textbf{3.73} &\textbf{1.47} &\textbf{3.85}&5.49&\textbf{3.44}&\textbf{2.04}&\textbf{3.56}&5.09&\textbf{3.18}&\textbf{1.91}\\ 

\midrule

\multirow{4}{*}{NeuMF}&
\multirow{2}{*}{NDCG} &
DP-SGD &11.22 &16.27 &10.21 &5.05 &10.65&15.25&9.24&7.03&10.18&15.28 &8.90&6.39\\
&& DP-Fair &\textbf{11.75} &15.51 &\textbf{10.98} &\textbf{4.52} &\textbf{11.09}&14.80&\textbf{9.98}&\textbf{4.82}&\textbf{10.47} &14.42&\textbf{9.48}&\textbf{4.94}\\
&\multirow{2}{*}{F1}&
DP-SGD &4.51 &6.78 &3.98 &2.79 &4.25&6.66&3.61&3.05&3.99&6.44&3.37&3.07\\
&& DP-Fair &\textbf{4.54} &6.31 &\textbf{4.10} &\textbf{2.21} &\textbf{4.27}&6.24&\textbf{3.78}&\textbf{2.46}&\textbf{4.03}&6.21&\textbf{3.49}&\textbf{2.72}\\ 

 \bottomrule
\end{tabularx}
\caption{\small The results of recommendation performance. The evaluation metrics are calculated based on top-10 predictions. The results are reported in percentage (\%) and the arrow indicate the favorable direction. Our best results are highlighted in bold. \label{tab:main}}
\vspace{-20pt}
\end{table*}

\subsection{Experimental Analysis}
\subsubsection{\bf Main Results} 
Based on Table \ref{tab:main}, several conclusions can be drawn. Firstly, the results are reported for three different privacy budget settings: the non-private setting ($\epsilon=\infty$), a loose privacy budget setting ($\epsilon=10$), and a tight budget setting ($\epsilon=1$). As expected and consistent with the literature, the utility in terms of NDCG and F1 degrades as the privacy constraint is tightened.
Secondly, our proposed DP-Fair algorithm outperforms DP-SGD in terms of overall utility for both NDCG and F1, regardless of the privacy budget $\epsilon$. This improvement is attributed to our custom noise addition and gradient clipping technique, which is also applicable in the non-private setting where the algorithm is identical to SGD with a re-ranking step. While it may seem counter-intuitive that enforcing fairness would improve utility, our results show that this is due to the improvement in the utility of inactive users, who make up 80\% of all users, thereby boosting the overall performance.
Finally, we observe a significant reduction in the $\mathcal{F}_{U}$ gap by DP-Fair over DP-SGD. This can be attributed to the post-processing step in DP-Fair, which identifies $k=10$ fairness-aware items out of the $K=20$ list.


\subsubsection{\bf Hyperparameter Effects.} 
In this experiment, we fix the privacy budget $\epsilon$ to 1. Our first objective is to examine the selection of the clipping bound. To simplify the analysis, we impose $C_u=C_v=C$. Based on the results presented in Figure \ref{fig:ablation}, we conclude that both excessively small or large values of $C$ have a negative impact on the NCDG. When the clipping parameter is too small, the average clipped gradient can be biased. Conversely, increasing the norm bound $C$ leads to the addition of more noise to the gradients. In addition, we investigate the impact of the fairness level selection. As the fairness requirements become stricter, the performance of the active group decreases, while that of the inactive group improves.


\section{Conclusions}
In this paper, we empirically observe the unfairness gap between the active and inactive users will be widened by the incorporation of DP-SGD in classical collaborative filtering based recommendation models. We propose custom differentially private gradient mapping incorporating an integer programming scheme to enhance its fairness between active and inactive users. Experiments on real-world e-commerce datasets show that DP-Fair outperforms DP-SGD in both utility and fairness metrics. 


\balance
\bibliographystyle{ACM-Reference-Format}
\bibliography{bibfile}

\end{document}